# RESEARCH OF OPEN SYSTEMS EVOLUTION BY THE METHOD OF MULTIFRACTAL ANALYSIS

E.N. Vertyagina

L.N. Gumilyov Eurasian National University, Astana, Kazakhstan, vert@inbox.ru

**Introduction**

Equilibrium processes in closed systems are well described by the research methods being already classical, which can also be used with some assumptions and simplifications for the analysis of processes in open systems, which are in the states near equilibrium. However, processes occurring in real open systems are far from equilibrium state and they can lead to synergetic effects, which are caused by coordinated behavior of system units. Traditional methods of analysis often just establish such behavior, but they are not able to reveal the mechanism of the process in full measure and to explain observed structural modifications from the standard positions.

At the contemporary stage of development, science requires a deeper explanation of causes of observed variations in open nonlinear systems, hence, more sensitive methods and techniques for a detailed description of the phenomena being studied. A qualitatively new approach to the research of nonlinear effects in complex systems were developed in the 2$^{nd}$ half of the last century, when I.Prigogine evolved the theory of self-organization [1] and B.Mandelbrot developed the theory of fractals [2]. From the positions of synergetics, a structural state, which appears in the system due to nonlinear processes, can be considered as a multifractal. Multifractal has a certain set of properties, which can be described by the power dependences and do not depend on the scale of the studying structure [3]. This hierarchy of multifractal structures allows transferring the characteristics and properties of the structure, being analyzed at micro- or nanoscale, upon the system as a whole. More over, synergetic parameters of the structure calculated at the multifractal analysis give additional information about occurring modifications and promote discovering of process mechanism.

**System parameters calculated by the method of multifractal analysis**

In order to calculate the multifractal characteristics with the purpose of quantitative description of structures there has been worked out computer software [4], which realizes the multifractal analysis method (MFA). The software allows to compute synergetic parameters of the studying system such as Rényi generalized fractal dimensions $D_q$ [5], multifractal spectrum function $f(a)$ [3], structural uniformity $R^2$ [6] and order $\Delta$ [7], information entropy [8-10] $S$ of the system under consideration and the chaotic structure being modeled from the same number of units as the studying structure.

To research the system by the MFA method one needs to analyze the pictures of structural distribution of system units. The computer code performs numeralization of the structure picture and treats the binary matrix, in which 1 corresponds to any structural formations, 0 is other supporting surface. The grid with changing cell size is put on the matrix; then the calculation of multifractal characteristics is performed according to the in work [3].

One of the basic parameters of a multifractal is probabilistic measure $p_i$, which shows relative fullness of cells covering the set.

Generalized dimensions $D_q$, also called as Rényi entropies [5], are determined by the following formulas:

$$D_q = \begin{cases} \dfrac{1}{1-q} \ln \sum_{i=1}^{N} p_i^q, & \text{if } q \neq 1 \\ -\sum_{i=1}^{N} p_i \ln p_i, & \text{if } q=1 \end{cases}, \qquad (1)$$

where $q_i$ is the index of singularity for the certain measure $p_i$. The values of $Dq$ are related with the multifractal spectrum functions $f(\alpha)$ by Legendre transformations. The physical meaning of $f(\alpha)$ function is that it represents Hausdorff dimension of a homogeneous fractal subset from the initial structure, which carries in a dominant weight contribution $p_i$ at given value of $q$.

$Dq$ and $f(\alpha)$ parameters give possibility to determine the degrees of uniformity and order of the studying structure. The uniformity parameter can be characterized by $p_i$ probabilities of filling of geometrically equivalent parts of the structure under consideration [6]. Uniform system is characterized by the parabolic function of multifractal spectrum $f(\alpha)$. In this case the square approximation of $f(\alpha)$ function by the least-squares method has the factor of reliability $R^2 = 1$. The higher degree of heterogeneity in the system corresponds to the more deviation of $f(\alpha)$ spectrum from parabola, i.e. $0 < R^2 < 1$. Hence, the $R^2$ value can be used as the degree of structure uniformity.

The order degree $\Delta$ can be determined as $D_1 - D_\infty$ [7], where information dimension $D_1$ characterizes the degree of structure symmetry breakdown, $D_\infty$ value is related with an error probability in the parameter determining. Then the more value of $\Delta = D_1 - D_\infty$ corresponds to the higher order degree of the system.

Entropy is the measure of uncertainty degree at the indicated level of the statistic description of the studying system. In a qualitative sense, the more the value of entropy is the more the number of essentially different microstates can be occupied by the object at the given macrostate. Shannon entropy [8] is a measure of information, which is needed for determining of the system site in a certain state:

$$S = -\sum_i p_i \ln p_i, \qquad (2)$$

where $p_i$ is the possibility of the system to be found in $i^{th}$ state. Haken's definition [9] of information entropy $S$ is based on the synergetic statement of $p_i$ measure for evolving systems. Authors [10] have shown that the physical meaning for parameterization of self-organization process was intrinsic for the specific value of information entropy, which is defined by the ratio $S$ to the number of grid cells, which contain the measure.

**MFA application for research of certain systems**
**1. Steel structure evolution under the action of laser radiation**

Samples made from high carbon alloyed steel have been preliminary treated and laser exposed; then they have been tested into depth by the traditionally used methods such as metallographic and X-ray structure analyses [6]. The results of the tests were compared with the parameters calculated by the "Multifractal analysis" software for the same microsections. The investigations have shown that the change of structure microhardness into samples depth correlates well to the variation of information entropy; dislocation density is a control parameter of self-organization process, it relates to the value of generalized fractal dimension. MFA allows getting additional information about the mechanisms of dissipation of laser radiation energy and about the generation of grain microstructures in the interface layer of steel.

Literature data also confirm that similar phenomena are observed in the structural materials under the action of irradiation [11], plasma [12], impact or plastic deformation [13]. Therefore, material behavior at the external influence is quite predictable, thus a lot of experimental tests can be replaced partly or maybe completely by the researches of microstructures pictures with the help of multifractal analysis.

## 2. Kinetics of luminescence decay at different temperatures

Experiments show that the process of electron excitation energy transfer between the molecules of donor and acceptor has a complex character; kinetic curves of luminescence decay are not exponential [14]. It seems to be caused by the formation of local structural clusters in the system and their evolution. The modeling of deactivation process and the following multifractal analysis of the structural distribution of interacting particles allows detecting the distinction of system fractal parameters at different temperatures of kinetic process [15]. The comparison of experimental dependencies and theoretical calculations shows that in the studying system the variation of kinetic characteristics of luminescence process at different matrix temperatures is related with reallocation of structural units of the system that leads to changing of the order parameter and the fractal dimension of distribution of reagents molecules. This complication of the dynamic structure can indicate the synergetic phenomena, which occur in it and promote more efficient photoprocesses.

## 3. Accumulation and annealing of luminescent centers in crystals

Modeling of radiation defect accumulation in a crystal shows that the aggregation sets consisting of the same type centers are formed in the system [16] at continuous radiation. Multifractal analysis of structural distribution of electron-hole centers reveals that at saturation of defect concentration there is a decrease of information entropy of the system. The defects structure self-organizes into a stable dynamic multifractal, which provides the dissipation of input energy by effective reallocation of its structural units. This process can go on infinitely long while the parameters of external influence do not change; but even in this case the structure tends to keep its own fractal parameters by searching for new ways of input energy outflow [17].

Simulation shows that multifractal structure can be generated not only under the irradiation, but also at the annealing of chaotic structure, which was formed accidentally [17]. Herewith the efficiency of thermoluminescence intensity allows determining the kind of the initial state of the system: multifractal or disorderly chaotic.

**Conclusion**

Thus, the multifractal approach can be applied for the characterization of different nature structures, e.g. distribution of electron-hole centers formed in natural minerals under the influence of spontaneous irradiation, nonuniformity of different materials surfaces (natural or formed due to technologic treatment). The application of MFA at different time moments of structure formation or destruction allows to reveal the dynamics of structural changes and to describe the system evolution as a whole.